\documentclass[12pt, preprint]{emulateapj}

\newcommand\arcpt{${{\lower3pt\hbox{$^{\prime\prime}$}}\atop{\raise4pt\hbox{.}}}$}
\newcommand\msun{$M_\odot$}
\newcommand\arcyear{$\hbox{{$^{\prime\prime}$} yr{$^{-1}$}}$}
\usepackage{lscape}                
\usepackage{longtable}             

\slugcomment{to appear in the {\it Astronomical Journal}}

\shorttitle{The Solar Neighborhood.\ XX.}
\shortauthors{Subasavage et al.}

\begin{document}

\title{The Solar Neighborhood. XX. Discovery and Characterization
of 21 New Nearby White Dwarf Systems}

\author{John P. Subasavage\altaffilmark{1}, Todd
J. Henry\altaffilmark{1}}

\affil{Department of Physics and Astronomy, Georgia State University,
Atlanta, GA 30302-4106}

\email{subasavage@chara.gsu.edu} 
\email{thenry@chara.gsu.edu}

\author{P. Bergeron}

\affil{D\'{e}partement de Physique, Universit\'{e} de Montr\'{e}al,
C.P. 6128, Succ. Centre-Ville, Montr\'{e}al, Qu\'{e}bec H3C 3J7,
Canada}

\email{bergeron@astro.umontreal.ca}

\author{P. Dufour}

\affil{Department of Astronomy and Steward Observatory, University of
Arizona, 933 North Cherry Avenue, Tucson, Arizona 85721, USA}

\email{dufourpa@as.arizona.edu}

\and

\author{Nigel C. Hambly}

\affil{Scottish Universities Physics Alliance (SUPA), Institute for
Astronomy, University of Edinburgh, Royal Observatory, Blackford Hill,
Edinburgh EH9 3HJ, Scotland, UK}

\email{nch@roe.ac.uk}

\altaffiltext{1}{Visiting astronomer, Cerro Tololo Inter-American
Observatory, National Optical Astronomy Observatory, which are
operated by the Association of Universities for Research in Astronomy,
under contract with the National Science Foundation.}


\begin{abstract}

We present medium resolution spectroscopy and multi-epoch $VRI$
photometry for 21 new nearby ($<$ 50 pc) white dwarf systems brighter
than $V \sim$ 17.  Of the new systems, ten are DA (including a wide
double degenerate system with two DA components), eight are DC, two
are DZ, and one is DB.  In addition, we include multi-epoch $VRI$
photometry for eleven known white dwarf systems that do not have
trigonometric parallax determinations.  Using model atmospheres
relevant for various types of white dwarfs (depending on spectral
signatures), we perform spectral energy distribution modeling by
combining the optical photometry with the near-infrared $JHK_S$ from
the Two Micron All-Sky Survey to derive physical parameters (i.e.,
$T_{\rm eff}$ and distance estimates).  We find that twelve new and
six known white dwarf systems are estimated to be within the NStars
and Catalog of Nearby Stars horizons of 25 pc.  Coupled with identical
analyses of the 56 white dwarf systems presented in Paper XIX of this
series, a total of 20 new white dwarf systems and 18 known white dwarf
systems are estimated to be within 25 pc.  These 38 systems of the 88
total studied represent a potential 34\% increase in the 25 pc white
dwarf population (currently known to consist of 110 systems with
trigonometric parallaxes of varying qualities).  We continue an
ongoing effort via CTIOPI to measure trigonometric parallaxes for the
systems estimated to be within 25 pc to confirm proximity and further
fill the incompleteness gap in the local white dwarf population.
Another 38 systems (both new and known) are estimated to be between 25
and 50 pc and are viable candidates for ground-based parallax efforts
wishing to broaden the horizon of interest.

\end{abstract}


\section{Introduction}

Given that all stars less than $\sim$8 \msun\ will eventually become
white dwarfs (WDs), the study of this class of objects is vital to
understanding our Galaxy. In particular, WD research addresses
questions concerning stellar structure and evolution, Galactic
components (i.e., thin disk, thick disk, halo), and even dark matter.
The oldest (i.e., coolest and least luminous) WDs help to constrain
the age of the Galactic components (particularly the thick disk) but
shine feebly and are only visible and easily characterized if they are
in our solar neighborhood.  A complete volume-limited sample provides
accurate statistics that can be reliably applied to the rest of the
Galaxy to infer the WD mass fraction, contribution to the halo
population, and WD number density within the disk.  Interesting
candidates can be targeted and more thoroughly scrutinized to identify
unusual and astrophysically compelling systems.

In a continuing effort to further complete the nearby WD sample, we
present spectra, optical and near-infrared photometry, as well as
modeled physical parameters for 21 new WD systems in the southern
hemisphere brighter than $V \sim$ 17.  Of these, 12 are estimated to
be within 25 pc, the horizon of the Catalog of Nearby Stars
\citep[CNS,][]{1991adc..rept.....G} and the NStars Database
\citep{2003fst3.book..111H}.  In addition, six previously known WDs
without trigonometric parallaxes are estimated to be within 25 pc.



\section {Candidate Selection}
\label{sec:candidates}

We conducted a southern hemisphere proper motion search using
digitized scans of the SuperCOSMOS Sky Survey (SSS), adopting a faint
magnitude limit of $R_{\rm 59F} =$ 16.5, called the SuperCOSMOS-RECONS
(SCR) survey.  The first wave of the survey adopted a lower proper
motion limit of 0.40\arcyear~with new discoveries published in
\citet{2004AJ....128..437H}, \citet{2004AJ....128.2460H}, and
\citet{2005AJ....129..413S, 2005AJ....130.1658S}.  A second effort
conducted by \citet{2007AJ....133.2898F} surveyed objects with proper
motions from 0.18-0.40\arcyear~and covered declinations $-$47$^\circ$
to $-$90$^\circ$.  The WD candidate selection effort also adopted the
lower proper motion limit of 0.18\arcyear, yet reached farther north
to declination $=$ 00$^\circ$.  Thus, this effort covered 92\% of the
southern sky, avoiding a few regions near the Galactic plane and the
Magellanic Clouds \citep[see Figure 1 of][]{2005AJ....130.1658S}.  A
full discussion of the search methodology can be found in
\citet{2004AJ....128..437H}.  Briefly, each of the four plate scans
used ($B_J, R_{\rm ESO}, R_{\rm 59F}, I_{\rm IVN}$) were positionally
mapped to a common coordinate system.  Any object that appeared on all
four plates and had a proper motion less than the lower limit was
discarded.  Objects that remained were then searched out to a radius
defined by a proper motion of 10.00\arcyear~and over 360$^\circ$ for
any other unpaired objects.  After a series of automated sifts for
false pairings, the remaining objects were cross-referenced with
previous proper motion catalogs to identify new and known objects.

\begin{figure}
\includegraphics[angle=90,width=85mm]{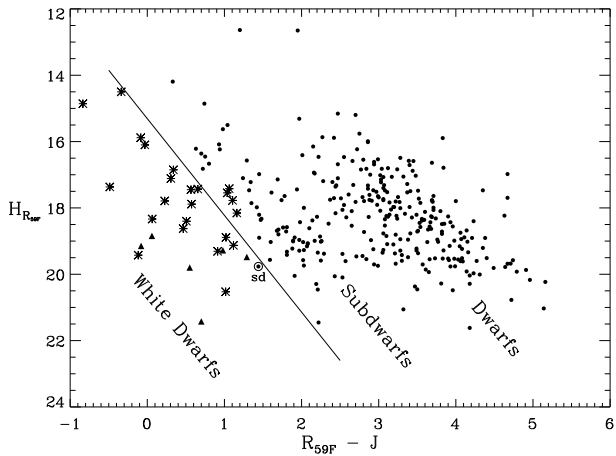}
\figcaption[redpromo.eps]{Reduced proper motion diagram used to select
WD candidates for spectroscopic follow-up.  Plotted are the 306 new
high proper motion objects from \citet{2005AJ....129..413S,
2005AJ....130.1658S} for which both $R_{\rm 59F}$ and $J$ magnitudes
were available.  The line is a somewhat arbitrary boundary between the
WDs (below) and the subdwarfs (just above).  Main sequence dwarfs fall
above and to the right of the subdwarfs, although there is significant
overlap.  Asterisks indicate the 23 new WDs reported here.  The six
filled triangles in the WD region are confirmed WDs from the SCR
survey and were published in \citet{2007AJ....134..252S}.  The
encircled point labeled ``sd'' is a confirmed subdwarf contaminant in
the WD region.
\label{rpm}}
\end{figure}

Near-infrared $JHK_S$ magnitudes were extracted from 2MASS for all
real objects (new and known) and a reduced proper motion (RPM) diagram
using the $R_{\rm 59F} - J$ color was generated (see Figure
\ref{rpm}).  RPM (in this case, designated by $H_{R_{\rm 59F}}$
because the $R_{\rm 59F}$ magnitude was used) is a quantity similar to
absolute magnitude and is defined by
\begin{equation}
H_{R_{\rm 59F}} = R_{\rm 59F} + 5 + 5 \log (\mu)
\end{equation}
where $\mu$ is proper motion.  It serves to relate two observed
quantities, apparent magnitude and proper motion, with two intrinsic
quantities, luminosity (absolute magnitude) and tangential velocity.
The advantages of using the $R_{\rm 59F} - J$ color is that all SCR
detections have an $R_{\rm 59F}$ magnitude (defined by the survey).
Given that WDs are relatively blue, the likelihood of registering a
near-infrared magnitude in the 2MASS database is greatest at $J$
because its limiting magnitude is $\sim$1.0 mags fainter than $K_S$.
Also, this color incorporates both optical plate and near-infrared
magnitudes that minimize the intrinsic uncertainties with plate
magnitudes (e.g., non-linearity).  Both new SCR discoveries and known
recoveries fell within the WD region of the RPM diagram.  Most of the
known objects had already been classified as WDs by previous
researchers.  Yet, a significant number of known high proper motion
(HPM) objects within this region were not classified as WDs.  It is
likely that they escaped identification because of the poor plate
magnitudes previously available.  For instance, all plate magnitudes
fainter than $m_{\rm pg} =$ 10.0 in the classic Luyten Half-Second
(LHS) Catalogue \citep{1979lccs.book.....L} and the New Luyten
Two-Tenths (NLTT) Catalogue \citep{nltt} are by-eye estimates
determined by \citet{1979lccs.book.....L}.  Rigorous calibrations
performed by the SSS team yield high quality plate magnitudes that
reveal a number of new, relatively bright, WDs.

\begin{figure}
\plotone{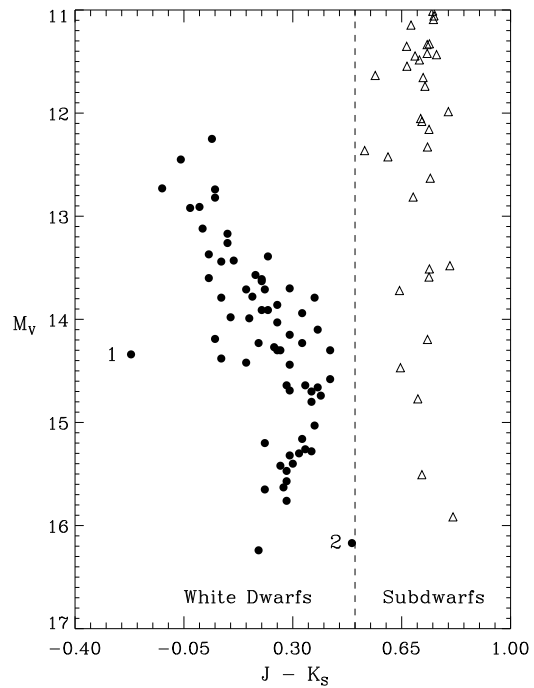} \figcaption[mvjk.eps]{Plot of infrared $J - K_S$
color \citep[transformed from the CIT system to the 2MASS system using
the transformation equations of][]{2001AJ....121.2851C} vs.~$M_V$ for
WDs within 25 pc \citep{2001ApJS..133..413B} and late-type subdwarfs
within 60 pc (W.-C. Jao 2008, private communication).  Note how this
particular color is less than 0.5 ({\it vertical dashed line}) over
all $M_V$ and is degenerate for the WDs.  Numbered points are
discussed in the text.
\label{mvjk}}
\end{figure}

As an additional constraint, especially for objects near the arbitrary
boundary that delineates the subdwarfs and the WDs, a $J - K_S$ color
sift was implemented.  As is evident in Figure \ref{mvjk}, this color
is degenerate for WDs (i.e., there is no unique absolute magnitude for
a given color) and is always less than 0.5 mag (for single WDs).  In
contrast, subdwarfs near the boundary typically have $J - K_S \sim$
0.6 or larger and thus there exists a significant color gap to
distinguish between the two luminosity classes.  Indeed, four new WDs
were spectroscopically identified that are found in the subdwarf
region of the RPM diagram.  Two exceptional points in the WD region
are: (1) WD 0548$-$001, which is a DQ WD \citep[discussed
in][]{dufour05} that has a magnetic field of $\sim$10 MG
\citep{2007ApJ...654..499K}; (2) WD 2251$-$070, which is a DZ WD
\citep[discussed in][]{dufour07} that is heavily line blanketed
\citep[][ Figure 11]{1993PASP..105..761W}.

Before the candidates were targeted for spectroscopic follow-up
observations, the basic linear plate relation of
\citet{2001Sci...292..698O} was used to estimate distances to
candidates, assuming they were WDs.  Only those candidates whose
distance estimates were within $\sim$30 pc were selected for
spectroscopic observations \citep[this step was omitted from the
procedures resulting in the new WD discoveries presented in][
hereafter referred to as Paper XIX]{2007AJ....134..252S}.  This
constraint favored cooler, nearby stars so that even though the sample
size of new systems presented here is smaller than that of Paper XIX
(21 vs.~33), more systems are estimated to be within 25 pc (twelve
vs.~eight). A total of 21 new WD systems (containing 22 WDs) were
spectroscopically confirmed and are hereafter referred to as the ``new
sample''.  Optical photometry observations were obtained to improve
distance estimates (discussed in $\S$ \ref{subsec:SED}).

Eleven known WDs without trigonometric parallaxes (hereafter referred
to as the ``known sample'') were also targeted for optical photometry
observations to improve their distance estimates.  The known sample
includes nine objects that previous authors predicted to be nearby
\citep[i.e.,][]{1993AJ....105.1033A, 2002ApJ...571..512H,
2008AJ....135.1225H, 2004AJ....127.1702K, 2007ApJ...654..499K,
2006AA...447..173P,2003ApJ...586L..95V} as well as two objects found
via RPM that were known WDs but whose distances had not yet been
estimated.

\section {Data and Observations}

\subsection {Astrometry and Nomenclature}

In keeping with the traditional naming convention for WDs, which uses
the object's epoch 1950 equinox 1950 coordinates, 2MASS coordinates
(when available, otherwise SSS coordinates were used) for the new
systems presented here were extracted and adjusted for proper motion
from the epoch of observation to epoch 2000 equinox 2000.  The
coordinates were then precessed to equinox 1950 using the IRAF task
{\it precess} and again adjusted for proper motion (opposite in
direction) to yield epoch 1950 equinox 1950 coordinates.

Proper motions for both the new and known samples were taken primarily
from the SCR proper motion survey, but in a few cases from elsewhere
in the literature.  The Appendix contains the proper motions used for
coordinate adjustment, as well as J2000.0 coordinates and alternate
names.

\subsection {Spectroscopy}

\begin{figure}[!t]
\plotone{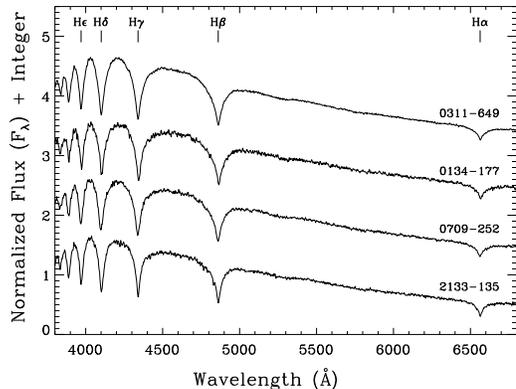}
\figcaption[dahot.eps]{Spectral plots of the hot ($T_{\rm eff}$ $\ge$
10000 K) DA WDs from the new sample, plotted in descending $T_{\rm
eff}$ as derived from the SED fits to the photometry.  Spectra have
been normalized at 5200 \AA~and offset by integer amounts.  The
inclusion of the spectrum of known object WD 2133$-$135 in this plot
is discussed in $\S$ \ref{subsec:individual}.
\label{dahot}}
\end{figure}

Spectroscopic observations were taken during two runs in May and
December 2006 at the Cerro Tololo Inter-American Observatory (CTIO)
1.5 m telescope as part of the Small and Moderate Aperture Research
Telescope System (SMARTS) Consortium.  The Ritchey-Chr{\'e}tien
spectrograph and Loral 1200 $\times$ 800 CCD detecter were used with
grating 09 (in first order), providing 8.6 \AA~resolution and
wavelength coverage from 3500 to 6900 \AA.  A slit width of 6$\arcsec$
was used and prevented the preferential light loss, at either the blue
or the red end, encountered in the data presented in Paper XIX.
Observations consisted of two exposures (typically 20-30 minutes each)
to permit cosmic ray rejection, followed by a comparison HeAr lamp
exposure to calibrate wavelength for each object.  Bias subtraction,
dome/sky flat-fielding, and extraction of spectra were performed using
standard IRAF packages.

Spectroscopic flux standards (two or three) were observed each night
to calibrate the response of the CCD across the dispersion axis.
However, because the selected flux standards are bright, neutral
density filters of either 2.5 or 5.0 magnitudes extinction were used.
Thus, the resulting science spectra are relatively flux calibrated
(i.e., the slopes are correct) but not on an absolute flux scale
(i.e., erg cm$^{-2}$ s$^{-1}$ \AA$^{-1}$).  All spectra are normalized
at 5200 \AA~for the sake of plotting.

\begin{figure}
\plotone{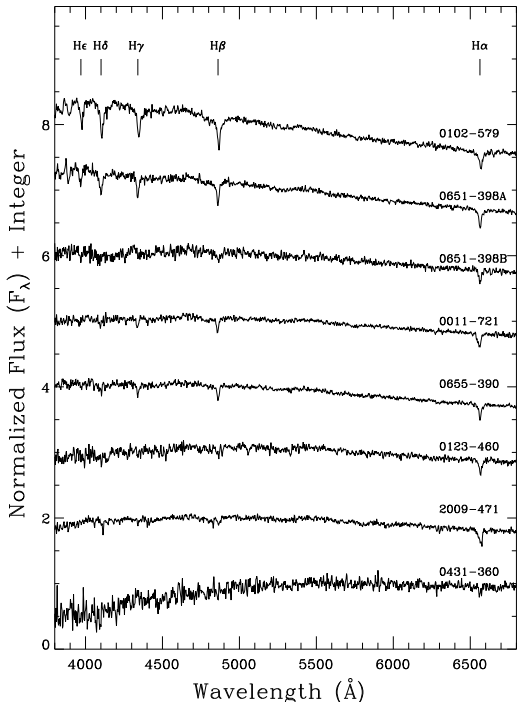}
\figcaption[dacool.eps]{Spectral plots of cool ($T_{\rm eff}$ $<$
10000 K) DA WDs from the new sample, plotted in descending $T_{\rm
eff}$ as derived from the SED fits to the photometry.  Spectra have
been normalized at 5200 \AA~and offset by integer amounts.  The
asymmetry in the H$\alpha$ feature in the spectrum of WD 2009$-$471 is
the result of a cosmic ray landing just redward of H$\alpha$ that
could not be reliably removed.
\label{dacool}}
\end{figure}

\begin{figure}[!t]
\plotone{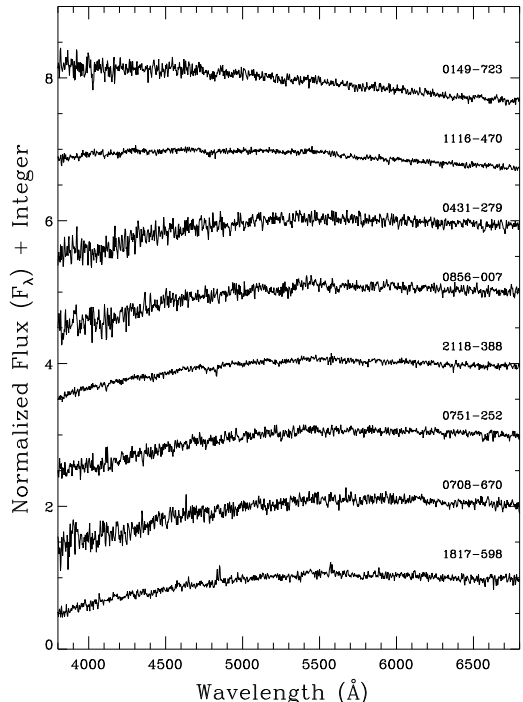}
\figcaption[dc.eps]{Spectral plots of featureless DC WDs from the new
sample, plotted in descending $T_{\rm eff}$ as derived from the SED
fits to the photometry.  Spectra have been normalized at 5200 \AA~and
offset by integer amounts.
\label{dc}}
\end{figure}

Spectra for the new DA WDs with $T_{\rm eff} \ge$ 10,000 K are plotted
in Figure \ref{dahot}, while spectra for new DA WDs with $T_{\rm eff}
<$ 10,000 K are plotted in Figure \ref{dacool}.  Spectra for the new
featureless DC WDs are plotted in Figure \ref{dc}.  Spectra, as well
as model fits, for two new calcium-rich DZs, the hottest DA, and a
helium-rich DB are shown in Figures \ref{dz} and \ref{dadb} and are
described in $\S$ \ref{subsec:SED}.

\subsection {Photometry}

Optical Johnson-Kron-Cousins $VRI$\footnote{The central wavelengths
for $V_J$, $R_{\rm KC}$, and $I_{\rm KC}$ are 5475, 6425, and 8075
\AA, respectively.} photometry for the new and known WDs was obtained
at the CTIO 0.9 m telescope during several observing runs from January
2006 through March 2008 as part of the SMARTS Consortium.  The 2048
$\times$ 2046 Tektronix CCD camera was used with the Tex 2 $VRI$
filter set.  Standard stars from \citet{1982PASP...94..244G} and
\citet{1992AJ....104..340L, 2007AJ....133.2502L} were taken nightly
through a range of airmasses to calibrate fluxes to the
Johnson-Kron-Cousins system and to calculate extinction corrections.
Bias subtraction and flat-fielding (using calibration frames taken
nightly) were performed using standard IRAF packages.  When possible,
an aperture of 14$\arcsec$ in diameter \citep[consistent
with][]{1992AJ....104..340L} was used to determine stellar flux.  If
cosmic rays fell within this aperture, they were removed before flux
extraction.  In cases of crowded fields or nearby companions, aperture
corrections were applied and ranged from 4$\arcsec$ to 12$\arcsec$ in
diameter using the largest aperture possible without including
contamination from neighboring sources.  Uncertainties in the optical
photometry are $\pm$ 0.03 mag in each band and incorporate both
internal (night-to-night variations) and external (fits to the
standard stars) uncertainties\footnote{A complete discussion of the
error analysis can be found in \citet{2004AJ....128.2460H}.}.  The
final optical magnitudes are listed in Table \ref{photometry}, as well
as the number of nights each object was observed.

\section {Analysis}
\subsection {Modeling of Physical Parameters}
\label{subsec:SED}

A detailed description of the model atmospheres used to model the WDs
can be found in Paper XIX and references within.  Briefly, the optical
$VRI$ and 2MASS near-infrared $JHK_S$ magnitudes are converted into
observed fluxes and compared to the resulting SEDs predicted by the
model atmospheres calculations.  The observed flux, $f_{\lambda}^m$,
is related to the model flux by the equation

\setcounter{footnote}{0}
\renewcommand{\thefootnote}{\alph{footnote}}
\begin{table*}[htp]
\centering
\begin{minipage}{170mm}
\tiny
\caption{Optical and Infrared
Photometry, and Derived Parameters for New and Known White Dwarfs.}
\begin{tabular}[c]{p{0.75in}rrrcrrrrrrr@{$\pm$}rlr@{$\pm$}lll}

\hline
\hline\\

\multicolumn{1}{c}{}                     &
\multicolumn{1}{c}{}                     &
\multicolumn{1}{c}{}                     &
\multicolumn{1}{c}{}                     &
\multicolumn{1}{c}{No.}                  &
\multicolumn{1}{c}{}                     &
\multicolumn{1}{c}{}                     &
\multicolumn{1}{c}{}                     &
\multicolumn{1}{c}{}                     &
\multicolumn{1}{c}{}                     &
\multicolumn{1}{c}{}                     &
\multicolumn{2}{c}{$T_{\rm eff}$}        & 
\multicolumn{1}{c}{}                     &
\multicolumn{2}{c}{Dist.}                &
\multicolumn{1}{c}{Spec.}                &
\multicolumn{1}{c}{}                     \\

\multicolumn{1}{c}{WD Name}              &
\multicolumn{1}{c}{$V_{\rm J}$}          &
\multicolumn{1}{c}{$R_{\rm KC}$}         &
\multicolumn{1}{c}{$I_{\rm KC}$}         &
\multicolumn{1}{c}{of Obs.}              &
\multicolumn{1}{c}{$J$}                  &
\multicolumn{1}{c}{$\sigma_J$}           &
\multicolumn{1}{c}{$H$}                  &
\multicolumn{1}{c}{$\sigma_H$}           &
\multicolumn{1}{c}{$K_S$}                &
\multicolumn{1}{c}{$\sigma_{K_S}$}       &
\multicolumn{2}{c}{(K)}                  &
\multicolumn{1}{c}{Comp.}                &
\multicolumn{2}{c}{(pc)}                 &
\multicolumn{1}{c}{Type}                 &
\multicolumn{1}{c}{Notes}                \\[4pt]

\hline\\

\vspace{0pt}\\[-10pt]
\multicolumn{18}{c}{New Spectroscopically Confirmed White Dwarfs} \\[4pt]
\hline
\vspace{0pt}\\[-3pt]

0011$-$721\dotfill       &  15.17   &  14.87   &  14.55   &   3   &  14.21  &  0.03  &  13.97  &  0.04  &  13.92  &  0.05  &   6439    &    152    &   H       &    17.8   &     2.9   &  DA8.0   & \tablenotemark{a}       \\%
0102$-$579\dotfill       &  16.35   &  16.17   &  15.98   &   2   &  15.67  &  0.07  &  15.57  &  0.16  &  15.76  &  Null  &   7866    &    375    &   H       &    44.4   &     7.5   &  DA6.5   & \tablenotemark{b}       \\%
0123$-$460\dotfill       &  16.30   &  15.94   &  15.57   &   3   &  15.11  &  0.04  &  14.84  &  0.06  &  14.91  &  0.10  &   5898    &    161    &   H       &    24.9   &     4.1   &  DA8.5   & \tablenotemark{a}       \\%
0134$-$177\dotfill       &  15.19   &  15.22   &  15.17   &   3   &  15.34  &  0.04  &  15.26  &  0.07  &  15.20  &  0.14  &  11329    &    560    &   H       &    47.7   &     8.4   &  DA4.5   & \tablenotemark{c}       \\
0149$-$723\dotfill       &  16.33   &  16.10   &  15.86   &   2   &  15.65  &  0.05  &  15.64  &  0.12  &  15.42  &  0.18  &   6972    &    298    &   He      &    36.2   &     5.8   &  DC      &                        \\%
0311$-$649\dotfill       &  13.27   &  13.34   &  13.36   &   2   &  13.45  &  0.02  &  13.46  &  0.03  &  13.57  &  0.05  &  11945    &    557    &   H       &    21.0   &     3.7   &  DA4.0   & \tablenotemark{d}       \\
0431$-$360\dotfill       &  17.03   &  16.55   &  16.08   &   2   &  15.48  &  0.06  &  15.17  &  0.08  &  15.23  &  0.18  &   5153    &    154    &   H       &    25.2   &     4.1   &  DA10.0  & \tablenotemark{a}       \\%
0431$-$279\dotfill       &  16.80   &  16.34   &  15.89   &   2   &  15.37  &  0.05  &  15.11  &  0.07  &  14.92  &  0.12  &   5330    &    146    &   H       &    24.7   &     4.0   &  DC      &                        \\%
0620$-$402\dotfill       &  16.20   &  15.89   &  15.60   &   2   &  15.27  &  0.04  &  15.13  &  0.09  &  15.24  &  0.17  &   5919    &    278    &   He(+Ca) &    25.3   &     4.0   &  DZ      &                        \\%
0651$-$398A\dotfill      &  15.46   &  15.23   &  14.98   &   2   &  14.71  &  0.04  &  14.55  &  0.05  &  14.49  &  0.11  &   7222    &    219    &   H       &    25.1   &     4.3   &  DA7.0   & \tablenotemark{e}       \\%
0651$-$398B\dotfill      &  16.07   &  15.76   &  15.44   &   2   &  15.10  &  0.05  &  14.90  &  0.08  &  14.71  &  0.13  &   6450    &    220    &   H       &    26.9   &     4.5   &  DA8.0   & \tablenotemark{a}       \\
0655$-$390\dotfill       &  15.11   &  14.81   &  14.48   &   3   &  14.15  &  0.03  &  13.88  &  0.04  &  13.89  &  0.07  &   6415    &    162    &   H       &    17.2   &     2.8   &  DA8.0   & \tablenotemark{a}       \\%
0708$-$670\dotfill       &  16.22   &  15.72   &  15.21   &   3   &  14.71  &  0.03  &  14.65  &  0.05  &  14.47  &  0.07  &   5108    &     74    &   He      &    17.5   &     2.7   &  DC      &                        \\%
0709$-$252\dotfill       &  14.38   &  14.36   &  14.34   &   2   &  14.39  &  0.03  &  14.39  &  0.04  &  14.49  &  0.09  &  10708    &    356    &   H       &    30.5   &     5.3   &  DA4.5   & \tablenotemark{f}       \\%
0751$-$252\dotfill       &  16.27   &  15.78   &  15.31   &   4   &  14.75  &  0.03  &  14.47  &  0.03  &  14.30  &  0.09  &   5159    &    107    &   H       &    17.8   &     2.9   &  DC      & \tablenotemark{g}       \\
0816$-$310\dotfill       &  15.43   &  15.21   &  15.05   &   3   &  14.92  &  0.04  &  14.73  &  0.07  &  14.83  &  0.12  &   6631    &    345    &   He(+Ca) &    23.8   &     3.8   &  DZ      &                        \\%
0856$-$007\dotfill       &  16.33   &  15.85   &  15.39   &   3   &  14.83  &  0.04  &  14.58  &  0.05  &  14.69  &  0.13  &   5309    &    126    &   H       &    19.3   &     3.2   &  DC      &                        \\%
1116$-$470\dotfill       &  15.52   &  15.20   &  14.86   &   3   &  14.45  &  0.03  &  14.37  &  0.06  &  14.35  &  0.09  &   5856    &    140    &   He      &    17.9   &     2.8   &  DC      &                        \\%
1817$-$598\dotfill       &  16.85   &  16.30   &  15.80   &   3   &  15.20  &  0.05  &  15.01  &  0.10  &  14.91  &  0.14  &   4960    &    145    &   H       &    20.9   &     3.4   &  DC      &                        \\%
1916$-$362\dotfill       &  13.60   &  13.69   &  13.77   &   2   &  14.10  &  0.03  &  14.22  &  0.04  &  14.21  &  0.07  &  24105    &   8797    &   He      &    41.7   &     7.7   &  DB      & \tablenotemark{h}       \\%
2009$-$471\dotfill       &  16.53   &  16.16   &  15.79   &   2   &  15.38  &  0.05  &  15.00  &  0.05  &  15.08  &  0.12  &   5827    &    162    &   H       &    27.0   &     4.5   &  DA8.5   & \tablenotemark{a}       \\%
2118$-$388\dotfill       &  16.55   &  16.09   &  15.65   &   2   &  15.16  &  0.04  &  14.92  &  0.07  &  15.05  &  0.12  &   5244    &    102    &   He      &    22.0   &     3.5   &  DC      &                        \\[4pt]%

\hline
\vspace{0pt}\\[-2pt]
\multicolumn{18}{c}{Known White Dwarfs (without Trigonometric Parallaxes)} \\[4pt]
\hline
\vspace{0pt}\\[-3pt]

0233$-$242\dotfill       &  15.94   &  15.43   &  14.93   &   3   &  14.45  &  0.03  &  14.34  &  0.05  &  14.12  &  0.07  &   5093    &     78    &   He      &    15.3   &     2.4   &  DC      & \tablenotemark{i}       \\
0707$-$320\dotfill       &  15.61   &  15.57   &  15.49   &   2   &  15.49  &  0.06  &  15.43  &  0.11  &  15.38  &  0.20  &   9900    &    440    &   H       &    47.8   &     8.2   &  DA5.0   &                        \\%
1223$-$659\dotfill       &  14.02   &  13.82   &  13.62   &   3   &  13.33  &  0.04  &  13.26  &  0.06  &  13.30  &  0.06  &   7690    &    220    &   H       &    14.5   &     2.5   &  DA6.5   & \tablenotemark{j}       \\
1241$-$798\dotfill       &  16.18   &  15.80   &  15.45   &   3   &  15.03  &  0.05  &  14.83  &  0.07  &  14.60  &  0.12  &   5618    &    143    &   He      &    22.1   &     3.5   &  DC/DQ   &                        \\
2007$-$219\dotfill       &  14.40   &  14.33   &  14.25   &   3   &  14.20  &  0.02  &  14.20  &  0.04  &  14.26  &  0.08  &   9556    &    242    &   H       &    25.7   &     4.4   &  DA5.5   & \tablenotemark{k}       \\
2133$-$135\dotfill       &  13.68   &  13.63   &  13.55   &   3   &  13.60  &  0.03  &  13.58  &  0.04  &  13.69  &  0.06  &  10182    &    281    &   H       &    20.4   &     3.5   &  DA5.0   & \tablenotemark{l}       \\%
2159$-$754\dotfill       &  15.04   &  14.92   &  14.80   &   2   &  14.72  &  0.04  &  14.67  &  0.07  &  14.55  &  0.10  &   8944    &    289    &   H       &    30.5   &     5.3   &  DA5.5   & \tablenotemark{m}       \\
2216$-$657\dotfill       &  14.55   &  14.47   &  14.41   &   2   &  14.54  &  0.04  &  14.50  &  0.06  &  14.53  &  0.09  &  10611    &    567    &   He      &    30.1   &     5.1   &  DZ      & \tablenotemark{n}       \\
2306$-$220\dotfill       &  13.75   &  13.83   &  13.89   &   2   &  14.13  &  0.03  &  14.18  &  0.05  &  14.33  &  0.07  &  16285    &   1273    &   H       &    33.3   &     6.1   &  DA3.0   & \tablenotemark{o}       \\
2336$-$079\dotfill       &  13.28   &  13.27   &  13.24   &   4   &  13.34  &  0.03  &  13.34  &  0.02  &  13.35  &  0.03  &  10946    &    304    &   H       &    18.9   &     3.3   &  DA4.5   & \tablenotemark{p}       \\
2351$-$335\dotfill       &  14.52   &  14.38   &  14.19   &   2   &  13.99  &  0.11  &  13.86  &  0.25  &  13.73  &  0.11  &   8068    &    401    &   H       &    20.1   &     3.5   &  DA6.0   & \tablenotemark{q}       \\[4pt]
\hline
\footnotetext[\tiny a]{\tiny Too cool for a reliable spectroscopic fit (i.e., minimal Balmer line absorption).}
\footnotetext[\tiny b]{\tiny Spectral fit yielded a $T_{\rm eff} =$ 8228 $\pm$ 138 K and log $g =$ 8.19 $\pm$ 0.12.  Common proper motion companion to NLTT 3566 (see $\S$ \ref{subsec:individual}).}
\footnotetext[\tiny c]{\tiny Spectral fit yielded a $T_{\rm eff} =$ 11613 $\pm$ 192 K and log $g =$ 8.14 $\pm$ 0.06.}
\footnotetext[\tiny d]{\tiny Spectral fit yielded a $T_{\rm eff} =$ 12632 $\pm$ 300 K and log $g =$ 7.63 $\pm$ 0.06 (see $\S$ \ref{subsec:individual}).}
\footnotetext[\tiny e]{\tiny Spectral fit yielded a $T_{\rm eff} =$ 7214 $\pm$ 135 K and log $g =$ 7.68 $\pm$ 0.19.  Common proper motion companion to WD 0651$-$398B (see $\S$ \ref{subsec:individual}).}
\footnotetext[\tiny f]{\tiny Spectral fit yielded a $T_{\rm eff} =$ 11208 $\pm$ 180 K and log $g =$ 8.12 $\pm$ 0.06.}
\footnotetext[\tiny g]{\tiny Common proper motion companion to LTT 2976 (see $\S$ \ref{subsec:individual}).}
\footnotetext[\tiny h]{\tiny Spectral fit yielded a $T_{\rm eff} =$ 27830 $\pm$ 1050 K and log $g =$ 7.97 $\pm$ 0.07 (see $\S$ \ref{subsec:individual} and Figure \ref{dadb}).}
\footnotetext[\tiny i]{\tiny \citet{2003ApJ...586L..95V} estimate a distance of 15 pc.}
\footnotetext[\tiny j]{\tiny \citet{2002ApJ...571..512H} estimate a distance of 10.79 pc.  \citet{2008AJ....135.1225H} estimate a distance of 12.05 pc.}
\footnotetext[\tiny k]{\tiny \citet{2002ApJ...571..512H} estimate a distance of 18.22 pc.}
\footnotetext[\tiny l]{\tiny Spectral fit yielded a $T_{\rm eff} =$ 10357 $\pm$ 158 K and log $g =$ 7.99 $\pm$ 0.07.  \citet{2006AA...447..173P} estimate a distance of 23.3 pc.}
\footnotetext[\tiny m]{\tiny \citet{2007ApJ...654..499K} estimate a distance of 14 pc.  \citet{2008AJ....135.1225H} estimate a distance of 14.24 pc.}
\footnotetext[\tiny n]{\tiny \citet{1993AJ....105.1033A} estimate a distance of 28 pc.}
\footnotetext[\tiny o]{\tiny \citet{2004AJ....127.1702K} estimate a distance of 33 pc.}
\footnotetext[\tiny p]{\tiny \citet{2002ApJ...571..512H} estimate a distance of 15.6 pc.  \citet{2008AJ....135.1225H} estimate a distance of 17.45 pc.}
\footnotetext[\tiny q]{\tiny \citet{2002ApJ...571..512H} estimate a distance of 12.41 pc.}
\end{tabular}
\label{photometry}\\[-4pt]
\end{minipage}
\normalsize
\end{table*}
\renewcommand{\thefootnote}{\arabic{footnote}}
\setcounter{footnote}{3}

\begin{equation}
f_{\lambda}^m= 4\pi~(R/D)^2~H_{\lambda}^m
\end{equation}
\noindent
where $R/D$ is the ratio of the radius of the star to its distance
from Earth, and $H_{\lambda}^m$ is the Eddington flux (dependent on
$T_{\rm eff}$, $\log g$, and atmospheric composition), properly
averaged over the corresponding filter bandpass.  We consider only
$T_{\rm eff}$ and the solid angle [$\pi(R/D)^2$] to be free
parameters, and the uncertainties of both parameters are obtained
directly from the covariance matrix of the fit. In this study, we
simply assume a value of $\log g=8.0$ for each star.  As discussed in
\citet{BRL, 2001ApJS..133..413B}, the main atmospheric constituent ---
hydrogen or helium --- is determined by comparing the fits obtained
with both compositions, or by the presence of H$\alpha$ in the optical
spectra.  The derived values for $T_{\rm eff}$ for each object are
listed in Table \ref{photometry}.  For the two new DZ stars,
spectroscopic modeling of metal lines as well as the photometry was
used to constrain $T_{\rm eff}$ (see below).Also listed are the
spectral types for each object determined based on their spectral
features.  The DAs have been assigned a half-integer temperature index
as defined by \citet{1999ApJS..121....1M}, where the temperature index
equals 50,400/$T_{\rm eff}$.

Our grids of model atmospheres and synthetic spectra for DA and DB
stars are described respectively in \citet{liebert2005} and
\citet{beauchamp96}.  The atmospheric parameters -- $T_{\rm eff}$,
log $g$ (and hydrogen abundance for the DB stars) -- are determined
from the optical spectra using the so-called spectroscopic technique
\citep[see, e.g.,][]{1992ApJ...394..228B}, which relies on a
detailed comparison between synthetic and observed normalized line
profiles.  The model spectra (convolved with a Gaussian instrumental
profile) and the optical spectrum are first normalized to a continuum
set to unity.  The calculation of $\chi^2$ is then carried out in
terms of these normalized line profiles only, and the best fitting
solution is obtained using the nonlinear least-squares method of
Levenberg-Marquardt \citep{pressetal92}, which is based on a steepest
descent method.  Spectroscopically derived values of $T_{\rm eff}$ and
log $g$ are given in the footnotes of Table \ref{photometry} for the
DA and DB stars.

\begin{figure}
\plotone{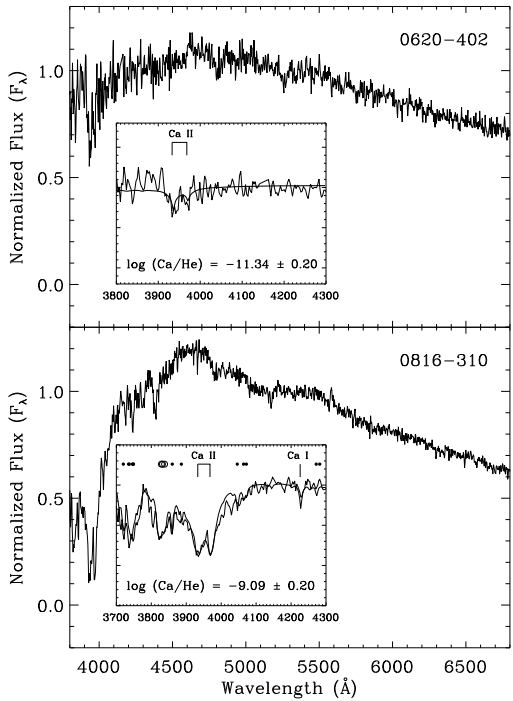}
\figcaption[dz.eps]{Spectral plots of the two DZ WDs from the new
sample.  The inset plots display the spectra ({\it thin lines}) in the
regions to which the models ({\it thick lines}) were fit.
Spectroscopic best fit physical parameters are listed below.
Prominent lines from Ca ({\it vertical lines}), Mg ({\it open
circles}), and Fe ({\it filled circles}) are labeled.  Both modeled
fits incorporated a slight hydrogen abundance of log (H/He) $=$ $-$3
(see $\S$ \ref{subsec:SED}).
\label{dz}}
\end{figure}

For the DZ stars' spectra in Figure \ref{dz}, we rely on the procedure
outlined in \citet{dufour07}.  We obtain a first estimate of the
atmospheric parameters by fitting the photometric SED with an assumed
value of the metal abundances (assuming solar abundance ratios).  We
then fit the optical spectrum to measure the metal abundances, and use
these values to improve our atmospheric parameters from the
photometric SED. This procedure is iterated until a self-consistent
photometric and spectroscopic solution is achieved.  In the cases of
the two DZ stars presented here, a slight abundance of hydrogen was
included [log (H/He) = $-$3] in the modeled fits.  As shown in
\citet{dufour07}, the Ca line widths and depths are affected by the
presence of hydrogen even if the abundance is not significant enough
to produce spectral signatures.  Our spectra were best fit with the
inclusion of hydrogen.  The values of $T_{\rm eff}$ listed in Table
\ref{photometry} for the new DZ stars were derived using this
additional spectroscopic modeling constraint [i.e., ``He($+$Ca)'' in
the composition column]; whereas, the known DZ (WD 2216$-$657) was
modeled using only photometry assuming a pure He atmosphere (i.e.,
``He'' in the composition column).

\begin{figure}
\plotone{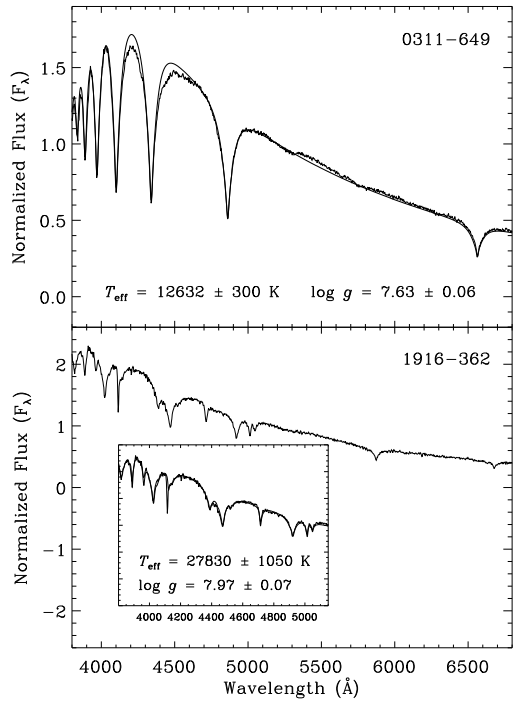} \figcaption[dadb.eps]{{\it Top}: Spectral plot of
the DA (hydrogen-rich) WD 0311$-$649 with the model fit ({\it thick
line}) overplotted.  Spectroscopic best fit physical parameters are
listed below. {\it Bottom}: Spectral plot of the DB (helium-rich) WD
1916$-$362.  The inset plot displays the spectrum ({\it thin line}) in
the region to which the model ({\it thick line}) was fit.
Spectroscopic best fit physical parameters are listed below.  Note
that the sharp absorption feature at $\sim$4100 \AA~is an artifact
produced by a cosmic ray that could not be reliably removed because of
its proximity to a true spectral feature.
\label{dadb}}
\end{figure}

Once the effective temperature and the atmospheric composition were
determined, we calculated the absolute visual magnitude using a
procedure identical to that outlined in Paper XIX \citep[i.e.,
combined the new photometric calibration of][~with evolutionary
models\footnote{See
http://www.astro.umontreal.ca/$\sim$bergeron/CoolingModels/.}]{holberg06}.
We then compared the absolute magnitude to the apparent $V$ magnitude
observed to derive a distance estimate for each star (reported in
Table \ref{photometry}).  Errors on the distance estimates incorporate
the errors of the photometry values as well as an error of 0.25 dex in
log {\it g}, which is the measured dispersion of the observed
distribution using spectroscopic determinations \citep[see Figure 9
of][]{1992ApJ...394..228B}.

\subsection {Comments on Individual Systems}
\label{subsec:individual}

{\bf WD 0102$-$579:} A new HPM object discovered during the SCR proper
motion survey ($\mu =$ 0.239\arcyear~at position angle 91.1$^\circ$,
\citealt{2007AJ....133.2898F}).  It is likely a common proper motion
companion to a previously known HPM object, the red dwarf NLTT 3566
($\mu =$ 0.257\arcyear~at position angle 87.5$^\circ$), separated by
110.6$\arcsec$ at position angle 206.6$^\circ$.  The {\it Hipparcos}
parallax for NLTT 3566 is 19.51 $\pm$ 3.13 mas \citep[distance = 51.3
$\pm$ 9.8 pc,][]{2007hnrr.book.....V} -- entirely consistent with our
distance estimate for the WD of 44.4 $\pm$ 7.5 pc.

{\bf WD 0311$-$649:} The hottest DA WD in the new sample ($T_{\rm eff}
=$ 11,945 $\pm$ 557 K).  The Balmer lines of our high S/N spectrum are
best fit to a spectral fitting model with $T_{\rm eff} =$ 12,632 $\pm$
300 K and log $g =$ 7.63 $\pm$ 0.06 (see Figure \ref{dadb}),
significantly less than the assumed value of log $g =$ 8.0 in the
photometric analysis.  This would imply that the object is low mass
($M =$ 0.42 $\pm$ 0.03 \msun) and more luminous thereby making it more
distant ($\sim$28 pc).  Given the age of our Galaxy, the lowest mass
WD that could have formed is $\sim$0.47
\msun~\citep{1984PhR...105..329I}.  If the mass is correct, it is
extremely unlikely that this object formed through single-star
evolution and thus is likely a multiple system.  Trigonometric
parallax measurements are underway to confirm the luminosity and hence
the mass via our Cerro Tololo Inter-American Observatory Parallax
Investigation \citep[CTIOPI,][]{2005AJ....129.1954J,
2005AJ....130..337C, 2006AJ....132.1234C, 2006AJ....132.2360H}.

{\bf WD 0620$-$402:} A new DZ WD whose modeled physical parameters
place this object in a lower temperature realm where high atmospheric
pressure effects exist that are not included in the Ca-rich (DZ)
models described in $\S$ \ref{subsec:SED}.  Thus, the model fit is
likely inaccurate for this object, so that the physical parameters
derived are not well constrained (especially given the low S/N of the
spectrum, see Figure \ref{dz}).

{\bf WD 0651$-$398AB:} A widely separated (87.2$\arcsec$ at position
angle 317.4$^\circ$) HPM double degenerate system discovered during
the SCR proper motion survey.  The A component ($\mu =$
0.229\arcyear~at position angle 344.7$^\circ$) has $T_{\rm eff} =$
7222 $\pm$ 219 K and the B component ($\mu =$ 0.227\arcyear~at
position angle 344.4$^\circ$) has $T_{\rm eff} =$ 6450 $\pm$ 220 K.  A
third object, WT 204, is separated by 59.3$\arcmin$ at position angle
67.2$^\circ$ and has a similar proper motion of 0.213\arcyear~at
position angle 341.9$^\circ$.  We have obtained spectra and optical
photometry, which indicate a spectral type of M3.0V and $V_J =$ 13.17,
$R_{\rm KC} =$ 12.03, $I_{\rm KC} =$ 10.60.  Using the main-sequence
distance relations of \citet{2004AJ....128.2460H}, we obtain a
distance estimate of 19.8 $\pm$ 3.1 pc, which is entirely consistent
with the distance estimates to the A and B components, 25.1 $\pm$ 4.3
pc and 26.9 $\pm$ 4.5 pc, respectively.  While highly unlikely that
this is a bound system (the third component's separation is $\sim$
90,000 AU), it is possible that they are part of a moving group.

{\bf WD 0751$-$252:} A new HPM object discovered during the SCR proper
motion survey ($\mu =$ 0.426\arcyear~at position angle 300.2$^\circ$,
\citealt{2005AJ....130.1658S}).  Its proper motion is similar to a
previously known HPM object, LTT 2976 ($\mu =$ 0.361\arcyear~at
position angle 303.8$^\circ$).  The distance estimate for the WD (17.8
$\pm$ 2.9 pc) is consistent with the {\it Hipparcos} parallax of 51.53
$\pm$ 1.46 mas (distance = 19.41 $\pm$ 0.57 pc,
\citealt{2007hnrr.book.....V}) for LTT 2976.  Thus, these two objects
likely form a system with a separation of 6.6$\arcmin$ ($\sim$ 8000
AU) at position angle 208.9$^\circ$.

{\bf WD 0816$-$310:} A new DZ WD whose spectrum has been reliably
reproduced using the methods appropriate for DZs described in $\S$
\ref{subsec:SED} (see Figure \ref{dz}).  The inset plot extends down
to 3700 \AA~(rather than 3800 \AA~for all other spectral plots) to
illustrate the validity of assuming solar abundance ratios (at least
for Ca, Mg, and Fe) given the quality of the fit.

{\bf WD 1916$-$362:} The only DB WD in the new sample.  The
photometric temperature ($T_{\rm eff} =$ 24,105 $\pm$ 8797 K) agrees
reasonably well with the spectroscopic temperature ($T_{\rm eff} =$
27,830 $\pm$ 1050 K) assuming a pure helium composition given that no
hydrogen features are visible in the spectrum.  At these temperatures,
trace amounts of hydrogen may be present without showing any spectral
signatures and would serve to reduce the spectroscopic temperature.
In fact, hydrogen would go unnoticed spectroscopically until at least
log (H/He) = $-$3.5 was included.  If we assume a log (H/He) = $-$4.5
(i.e., not spectrally visible), we obtain a spectroscopic solution
with $T_{\rm eff} =$ 25,800 K, slightly more consistent with the
photometric temperature.  It is possible that trace amounts of
hydrogen are present, however, without additional information (i.e., a
trigonometric parallax), we have chosen to adopt the spectroscopic
result assuming a pure helium composition.

We note that the photometric temperature is far more uncertain.  The
optical photometry used to fit the SED is rather insensitive to
effective temperature because the magnitudes fall largely on the
Rayleigh-Jeans tail of the SED for such a hot object.  Therefore, the
spectroscopically determined effective temperature is significantly
better constrained (spectrum and fit are plotted in Figure
\ref{dadb}).


{\bf WD 2133$-$135:} A DA WD that was uncovered during our sift of the
SSS database.  It is also included in \citet{2006AA...447..173P}
(labeled HE 2133$-$1332 in that work), in which they analyzed spectra
of 398 DA WDs to determine kinematic properties of the sample as part
of the SN Ia Progenitor surveY (SPY).  The authors state that their
sample was generated from databases of known WDs; however, we
performed a thorough search for a previous spectroscopic confirmation
of this object's WD status and were unsuccessful\footnote{Resources
searched include the WD database of \citet{1999ApJS..121....1M} (the
current Web-based catalog can be found at
http://heasarc.nasa.gov/W3Browse/all/mcksion.html) as well as a
previous WD publication using data from the Hamburg/ESO (HE) survey
\citep[i.e.,][]{2001A&A...366..898C}.}.  We have confirmed that this
object was selected to be observed by the SPY project using data taken
from the Hamburg/ESO (HE) survey in which there was no previous
confirmation spectrum but rather a high confidence in this object's
luminosity class (R. Napiwotzki 2008, private communication).  Thus,
\citet{2006AA...447..173P} is the first publication to confirm this
object's WD status (hence, it is a member of the known sample) yet
this publication is the first to present a confirmation spectrum (see
Figure \ref{dahot}).  The authors estimate a spectroscopic distance of
23.3 pc, consistent with our distance estimate of 20.4 $\pm$ 3.5 pc.

{\bf WD 2159$-$754:} A known DA WD that has been analyzed
spectroscopically by \citet{2007ApJ...654..499K} for which they
determine the spectroscopic $T_{\rm eff} =$ 9040 $\pm$ 80 K,
moderately consistent with our photometric $T_{\rm eff} =$ 9556 $\pm$
242 K.  In addition, they conclude that this object has a large
surface gravity (log $g =$ 8.95 $\pm$ 0.12) and hence a large mass
(1.17 $\pm$ 0.04 \msun).  Thus, their estimated distance (14 pc) is
significantly closer than our estimated distance (30.5 $\pm$ 5.3 pc)
based on an assumed log $g =$ 8.0.  Trigonometric parallax
measurements are underway to confirm its luminosity and hence its
mass.

\section {Discussion}

We continue to fill in the nearby WD sample with the discovery of 21
new WD systems brighter than $V \sim$ 17.  While the sample size is
smaller than the number of new WDs presented in Paper XIX, the number
of new systems estimated to be within 25 pc is larger (eight reported
in Paper XIX vs.~twelve reported here for a total of 20 new WDs), due
largely to the revised criteria set for candidates to be targeted for
follow-up spectroscopic observations (see $\S$ \ref{sec:candidates}).
In addition, of the known samples evaluated in this effort, a total of
18 WDs are estimated to be within 25 pc (twelve reported in Paper XIX
and six reported here).  Combining the new and known samples from both
publications, we have found a total of 38 WDs estimated to be within
25 pc -- a volume that currently contains 110 WD systems disregarding
any quality constraints on the parallaxes.\footnote{This sample of WDs
was compiled using parallaxes, both for the WDs as well as for
additional components if the WD is part of a multiple system, from the
Yale Parallax Catalog \citep{1995gcts.book.....V}, {\it Hipparcos}
\citep{1997A&A...323L..49P, 2007hnrr.book.....V}, and other recent
efforts involving trigonometric parallaxes
\citep[e.g.,][]{2003A&A...404..317S, 2004ApJS..150..455G,
2007A&A...470..387D}.  A comprehensive list, including weighted mean
parallaxes when multiple parallaxes are avaible for a system, is
included as an electronic supplement to this publication and can also
be found at http://www.DenseProject.com.}  Trigonometric parallax
determinations are underway to confirm membership to the 25 pc sample
but should the distance estimates prove accurate, the local sample of
WDs will be increased by 34\%.  Once these nearby WDs are confirmed,
focused efforts can yield crucial WD masses and permit companionship
assessments.

To ensure a reliable sample, we have begun to adopt the quality
constraint that the trigonometric parallax error cannot be greater
than 10\% of the parallax.  Given the precision of ground-based
trigonometric parallaxes ($\sim$2.0 mas or better), this constraint is
entirely reasonable (even for a system at the 25 pc horizon, the 10\%
constraint amounts to an error of 4.0 mas).  With the constraint in
place, thirteen systems (five in the north and eight in the south) are
eliminated from the known 25 pc WD sample (now with a total of 97
systems).  To better constrain their distances, we are currently
measuring trigonometric parallaxes for seven of these systems in the
southern hemisphere (one system, WD 1043$-$188 is too close to a
companion whose brightness exceeds the bright limit of the 0.9m at
CTIO).  We would encourage northern hemisphere parallax programs to do
the same for the five systems in the north (WD 0644$+$025, WD
0955$+$247, WD 1309$+$853, WD 1919$+$145, and WD 2117$+$539).

If we separate new and known samples from Paper XIX and this work by
proper motion, we see that the majority of added WD neighbors
estimated to be within 25 pc is found at lower proper motions.  Table
\ref{diststats} breaks down this complete sample into proper motion
bins of 0.2\arcyear~below 1.0\arcyear~(the first value in each column
corresponds to the systems presented in Paper XIX while the second
value corresponds to those systems reported here).  A quick summation
shows the number of systems with $\mu \ge$ 0.6\arcyear~(12) is dwarfed
by the number of systems with $\mu <$ 0.6\arcyear~(26), including five
that have $\mu <$ 0.2\arcyear~(of which, four have $\mu$ between
0.18-0.2\arcyear).  These results support the notion that a
significant number of nearby WDs may still be found at very low proper
motions.

Other recent efforts have focused on the local WD population.  In
particular, \citet{2008AJ....135.1225H} have targeted the 20 pc sample
determined by trigonometric parallaxes as well as
photometric/spectroscopic parallaxes.  The authors estimate the local
WD density based on the 13 pc WD sample using the assumption that the
13 pc sample is largely complete.  Paper XIX contained two objects (WD
0821$-$669 and WD 1202$-$232) previously unknown to be nearby, whose
distance estimates as well as unpublished trigonometric parallaxes
(Subasavage et al. 2008, in preparation) placed them within 13 pc.
These have been included by \citet{2008AJ....135.1225H} in their
updated estimate of the local WD density.  Given that none of the new
WD discoveries in this publication have distance estimates within 13
pc, the notion that the 13 pc WD sample is largely complete is
supported.  The sample of \citet{2008AJ....135.1225H} includes five
new WD discoveries from Paper XIX whose distance estimates lie within
20 pc (WD 0121$-$429, WD 0344$+$014, WD 0821$-$669, WD 2008$-$600, WD
2138$-$332) as well as one from this paper (WD 0751$-$252, see $\S$
\ref{subsec:individual}).  The authors estimate that the 20 pc sample
is $\sim$80 \% complete and that $\sim$33 $\pm$ 13 WDs remain to be
discovered between 13 and 20 pc (again, assuming the 13 pc sample is
complete).  Five systems reported here (not including WD 0751$-$252)
are estimated to lie within 20 pc.  Thus, we are filling the
incompleteness gap.

%
\begin{table}[t]
\caption{Distance Estimate Statistics for New and Known White Dwarf Systems$^{\rm a}$.} 
\begin{tabular}{p{1.5in}r@{ + }lr@{ + }lr@{ + }l}
\label{diststats}\\
\hline
\hline\\

           \multicolumn{1}{c}{Proper motion}&
           \multicolumn{2}{c}{d $\leq$ 10}&
           \multicolumn{2}{c}{10 $<$ d $\leq$ 25}&
           \multicolumn{2}{c}{d $>$ 25} \\

           \multicolumn{1}{c}{(arcsec yr$^{-1}$)}&
           \multicolumn{2}{c}{(pc)}&
           \multicolumn{2}{c}{(pc)}&
           \multicolumn{2}{c}{(pc)} \\[4pt]

\hline\\[-3pt]
%
%

$\mu$ $\geq$ 1.0\dotfill         &  1  &  0  &  6  &  0  &   1  &   0   \\
1.0 $>$ $\mu$ $\geq$ 0.8\dotfill &  0  &  0  &  0  &  1  &   0  &   0   \\
0.8 $>$ $\mu$ $\geq$ 0.6\dotfill &  0  &  0  &  2  &  2  &   2  &   1   \\
0.6 $>$ $\mu$ $\geq$ 0.4\dotfill &  0  &  0  &  6  &  4  &  11  &   2   \\
0.4 $>$ $\mu$ $\geq$ 0.2\dotfill &  0  &  0  &  4  &  7  &  20  &  11   \\
0.2 $>$ $\mu$ $\geq$ 0.0\dotfill &  0  &  0  &  1  &  4  &   2  &   0   \\
{~~~~}Total\dotfill              &  1  &  0  & 19  &  18 &  36  &  14   \\[4pt]

\hline\\[-2pt]
\parbox{250pt}{
$^{\rm a}$~~~ The first number is from Paper XIX, and the second
number is from this paper}
\end{tabular}
\end{table}

A star's mass is one of its most important characteristics that, when
coupled with composition and luminosity, defines several fundamental
properties of that star such as its internal structure, future
evolution, and total lifetime.  The same is true for WDs; however,
only four WDs have measured astrometric mass determinations better
than 5\% \citep[Sirius B, Procyon B, 40 Eri B, and Stein 2051
B,][]{1998ApJ...494..759P}.  The identification of WDs in binaries, in
particular, double degenerate systems, may increase the number of
accurate astrometric masses of WDs if the system were resolvable using
high-precision astrometric techniques (e.g., speckle, adaptive optics,
or interferometry via {\it Hubble Space Telescope}'s Fine Guidance
Sensors).  Of course, the nearer the system is to the Sun, the greater
its projected separation and more likely it is to be resolved, all
else being equal.  A sizable sample of accurate WD masses will help
constrain and revise WD models, while comprehensive searches for
companions in a volume-limited sample will allow an accurate
multiplicity fraction to be determined.

We continue to measure trigonometric parallaxes with good precision
from the ground ($\sim$1.1 milliarcsecond errors on average,
Subasavage et al. 2008, in preparation) for the nearby WD sample as
part of the CTIOPI program.  In addition, nearly all WDs within 15 pc
in the southern hemisphere (including those we have presented in Paper
XIX and here) have been targeted for long-term astrometric monitoring
in search of perturbations from unseen companions.  This effort is an
extension of the Astrometric Search for Planets Encircling Nearby
Stars \citep[ASPENS,][]{2003AAS...203.4207K} that targets all red and
white dwarfs within 10 pc in the southern hemisphere.  Because of the
spectral signatures of WDs (broad absorption lines or no lines at
all), astrometry is currently the only practical method for detecting
sub-stellar/planetary companions to WDs.  Unlike radial velocity
variations used to detect planetary systems, astrometric signatures
are linearly related to the distance to the system (the farther the
system, the smaller the astrometric signature).  Thus, a careful
evaluation of the nearest WDs provides the highest probability for
detecting astrometric signatures of companions to WDs.  

\section {Acknowledgments}

We wish to thank the referee of this manuscript for helpful comments
that improved the clarity of this publication.  The RECONS team at
Georgia State University wishes to thank the NSF (grant AST 05-07711),
NASA's Space Interferometry Mission, the Space Telescope Science
Institute (grant HST-GO-10928.01-A), and GSU for their continued
support of our study of nearby stars.  We also thank the members of
the SMARTS Consortium, who enable the operations of the small
telescopes at CTIO where all of the data in this work were collected.
J.~P.~S.~is indebted to Wei-Chun Jao for the use of his photometry
reduction pipeline.  P.~B.~is a Cottrell Scholar of Research
Corporation and would like to thank the NSERC Canada and the fund
FQRNT (Qu\'ebec) for their support.  N.~C.~H.~would like to thank
colleagues in the Wide Field Astronomy Unit at Edinburgh for their
efforts contributing to the SSS effort; particular thanks go to Mike
Read, Sue Tritton, and Harvey MacGillivray.  This work has made use of
the SIMBAD, VizieR, and Aladin databases, operated at the CDS in
Strasbourg, France.  We have also used data products from the Two
Micron All Sky Survey, which is a joint project of the University of
Massachusetts and the Infrared Processing and Analysis Center, funded
by NASA and NSF.

\section{Appendix}

In order to ensure correct cross-referencing of names for the new and
known WD systems presented here, Table \ref{alternate} lists
additional names found in the literature.  Objects for which there is
an NLTT designation also have the corresponding L or LP designation
found in the NLTT catalog.  These L or LP names are listed here
because the NLTT designations were not published in the original
catalog, but rather are the record numbers in the electronic version
of the catalog and have been adopted out of necessity.


\clearpage


\begin{longtable}[c]{p{0.75in}ccccll}
\caption{Astrometry and Alternate Designations for New and Known White Dwarfs.}
\label{alternate}\\
\hline
\hline\\

\multicolumn{1}{c}{WD Name}          & 
\multicolumn{1}{c}{RA}               & 
\multicolumn{1}{c}{Dec}              & 
\multicolumn{1}{c}{PM}               & 
\multicolumn{1}{c}{PA}               & 
\multicolumn{1}{c}{Ref}              & 
\multicolumn{1}{c}{Alternate Names}  \\

\multicolumn{1}{c}{}                 & 
\multicolumn{1}{c}{(J2000.0)}        & 
\multicolumn{1}{c}{(J2000.0)}        & 
\multicolumn{1}{c}{(arcsec yr$^{-1}$)}   & 
\multicolumn{1}{c}{(deg)}            & 
\multicolumn{1}{c}{}                 & 
\multicolumn{1}{c}{}                 \\[4pt]               

\hline\\

%

%

\vspace{0pt}\\[-12pt]
\multicolumn{7}{c}{New Spectroscopically Confirmed White Dwarfs} \\[4pt]
\hline
\vspace{0pt}\\[-5pt]

0011$-$721\dotfill       &  00 13 49.91  &  $-$71 49 54.2  &  0.326  &  141.3  &  S  & NLTT 681, LP 50-73                 \\
0102$-$579\dotfill       &  01 04 12.14  &  $-$57 42 48.6  &  0.239  &  091.1  &  S  & SCR 0104$-$5742                    \\
0123$-$460\dotfill       &  01 25 18.03  &  $-$45 45 31.1  &  0.759  &  137.8  &  S  & SCR 0125$-$4545                    \\
0134$-$177\dotfill       &  01 37 15.16  &  $-$17 27 22.6  &  0.319  &  189.3  &  S  & NLTT 5424, LP 768-192              \\
0149$-$723\dotfill       &  01 50 38.49  &  $-$72 07 16.7  &  0.334  &  223.9  &  S  & SCR 0150$-$7207                    \\
0311$-$649\dotfill       &  03 12 25.68  &  $-$64 44 10.8  &  0.190  &  105.6  &  S  & WT 106, LEHPM 1-3159               \\
0431$-$360\dotfill       &  04 32 55.87  &  $-$35 57 28.9  &  0.301  &  084.1  &  S  & LEHPM 2-1182                       \\
0431$-$279\dotfill       &  04 33 33.58  &  $-$27 53 24.8  &  0.403  &  092.4  &  S  & NLTT 13532, LP 890-39, LEHPM 2-405 \\
0620$-$402\dotfill       &  06 21 41.64  &  $-$40 16 18.7  &  0.379  &  166.0  &  S  & LEHPM 2-505                        \\
0651$-$398A\dotfill      &  06 53 30.21  &  $-$39 54 29.1  &  0.227  &  344.4  &  S  & WT 202                             \\
0651$-$398B\dotfill      &  06 53 35.34  &  $-$39 55 33.3  &  0.229  &  344.7  &  S  & WT 201                             \\
0655$-$390\dotfill       &  06 57 05.90  &  $-$39 09 35.7  &  0.340  &  242.6  &  S  & NLTT 17220, L 454-9, LTT 2692      \\
0708$-$670\dotfill       &  07 08 52.28  &  $-$67 06 31.4  &  0.246  &  246.3  &  S  & SCR 0708$-$6706                    \\
0709$-$252\dotfill       &  07 11 14.39  &  $-$25 18 15.0  &  0.223  &  334.4  &  S  & SCR 0711$-$2518                    \\
0751$-$252\dotfill       &  07 53 56.61  &  $-$25 24 01.4  &  0.426  &  300.2  &  S  & SCR 0753$-$2524                    \\
0816$-$310\dotfill       &  08 18 40.26  &  $-$31 10 20.3  &  0.842  &  162.6  &  S  & SCR 0818$-$3110                    \\
0856$-$007\dotfill       &  08 59 12.91  &  $-$00 58 42.9  &  0.202  &  125.8  &  S  & NLTT 20690, LP 606-32              \\
1116$-$470\dotfill       &  11 18 27.20  &  $-$47 21 57.0  &  0.322  &  275.1  &  S  & SCR 1118$-$4721                    \\
1817$-$598\dotfill       &  18 21 59.54  &  $-$59 51 48.5  &  0.365  &  194.9  &  S  & SCR 1821$-$5951                    \\
1916$-$362\dotfill       &  19 20 02.83  &  $-$36 11 02.7  &  0.208  &  132.0  &  S  & SCR 1920$-$3611                    \\
2009$-$471\dotfill       &  20 12 48.75  &  $-$46 59 02.5  &  0.244  &  136.3  &  S  & WT 689                             \\
2118$-$388\dotfill       &  21 22 05.59  &  $-$38 38 34.7  &  0.186  &  113.5  &  S  & SCR 2122$-$3838                    \\

\vspace{-0pt}\\[-5pt]
\hline
\vspace{-0pt}\\[-3pt]
\multicolumn{7}{c}{Known White Dwarfs (without Trigonometric Parallaxes)} \\[4pt]
\hline
\vspace{-05pt}\\[-2pt]

0233$-$242\dotfill       &  02 35 21.80  &  $-$24 00 47.3  &  0.620  &  189.5  &  S  & LHS 1421, NLTT 8435, LP 830-14   \\
0707$-$320\dotfill       &  07 09 25.07  &  $-$32 05 07.3  &  0.551  &  338.2  &  L  & LHS 1898, NLTT 17486, LP 896-18  \\
1223$-$659\dotfill       &  12 26 42.02  &  $-$66 12 18.5  &  0.190  &  182.0  &  L  & NLTT 30737, LP 104-2, GJ 2092    \\
1241$-$798\dotfill       &  12 44 52.70  &  $-$80 09 27.8  &  0.578  &  306.3  &  L  & LHS 2621, NLTT 31694, LP 38-80   \\
2007$-$219\dotfill       &  20 10 17.51  &  $-$21 46 45.6  &  0.311  &  158.0  &  L  & NLTT 48815, LP 870-43            \\
2133$-$135\dotfill       &  21 36 16.38  &  $-$13 18 34.5  &  0.297  &  120.2  &  S  & NLTT 51636, Ross 203, HE 2133-1332 \\
2159$-$754\dotfill       &  22 04 20.84  &  $-$75 13 26.1  &  0.523  &  277.6  &  S  & LHS 3752, NLTT 52728, LP 48-15   \\
2216$-$657\dotfill       &  22 19 48.31  &  $-$65 29 17.6  &  0.660  &  160.1  &  L  & LHS 3794, NLTT 53489, LP 119-34  \\
2306$-$220\dotfill       &  23 08 40.78  &  $-$21 44 59.6  &  0.350  &  109.0  &  S  & NLTT 55932, LP 877-69            \\
2336$-$079\dotfill       &  23 38 50.74  &  $-$07 41 19.9  &  0.034  &  126.6  &  Su & GD 1212                          \\
2351$-$335\dotfill       &  23 54 01.14  &  $-$33 16 30.3  &  0.500  &  216.5  &  L  & LHS 4040, NLTT 58330, LP 936-12  \\[3pt]

\hline\\[-22pt]
\end{longtable}
\begin{center}
\footnotesize
\parbox{420pt}{
References. ---
(L)  ~\citealt{1979lccs.book.....L, nltt}; 
(S) ~\citealt{2005AJ....129..413S, 2005AJ....130.1658S, 2007AJ....133.2898F}, this work;
(Su)~Subasavage et al. 2008, in preparation
}

\end{center}

\clearpage

\end{document}